\documentstyle[aps,manuscript,epsf]{revtex}
\begin{document}

\title{Remarks on the Theory of Cosmological Perturbation}
\author{Wenbin Lin
\\Institue of High Energy Physics, Chinese Academy of Science,
P.O. Box 918-4, Beijing 100039, P.R. China}

\maketitle

\begin{abstract}

It is shown that the power spectrum defined in the Synchronous Gauge 
can not be {\em directly} used to calculate the predictions of 
cosmological models on the large-scale structure of universe, which 
should be 
calculated {\em directly}
by a suitable gauge-invariant power spectrum or the power spectrum 
defined in the Newtonian Gauge.

\end{abstract}
~
~\\

\newpage

It is indispensable to adopt the linearized Einstein equation to solve 
the problem of the evolution of density perturbation in the expanding 
universe, and this involves the freedom of gauge. Currently, the 
density 
perturbation is widely calculated in the Synchronous Gauge ($SG$) frame 
and then {\em directly} compared with the observations of the 
large-scale structure of universe, e.g., the abundance and distribution 
of the 
galaxies or the clusters (e.g., Refs.
\cite{lss,bardeen1,white,dodelson,gawiser}). Because the power 
spectrums of density perturbation differ 
dramatically for small wavenumbers in different gauges, this raises an 
important problem: Does the density perturbation in the $SG$ frame 
characterize {\em directly} the large-scale structure of universe
which is relative to the unperturbed expanding background of universe?
I do not think so.

Because the density perturbations defined in different gauges are 
generally different from each other, we need to construct some suitable 
gauge-invariant
variables that can be directly confronted with the large-scale 
structure of universe. The gauge-invariant approach to the 
gravitational 
perturbation,
which was pioneered by Bardeen\cite{bardeen2}, has been applied to the 
cosmological perturbation (e.g., Refs.\cite{kodama1,mukhanov}). However, 
there exist some errors in the comparison between the theoretical 
calculations and the
observations of the large-scale structure of universe.

The most general form of the line element for a spatially flat 
background and scalar metric perturbations can be written as
\cite{mukhanov}
\begin{equation}
  ds^2=a^2\{(1+2\Psi)d\tau^2-2B_{\mid i}dx^id\tau-[\delta_{ij}
-2(\Phi\delta_{ij}-E_{\mid_{ij}})]dx^idx^j~\},
\end{equation}
where $a$ and $\tau$ are the conformal cosmic expansion scale factor 
and the conformal cosmic time; ``$_\mid$'' denotes the background 
three-dimensional covariant derivative. The corresponding perturbed 
energy-momentum tensor $T^{\mu}_{\nu}$ has the form 
\begin{eqnarray}
  \nonumber T^0_{~0}&=&\rho(1+\epsilon)~,\\
  \nonumber T_0^{~i}&=&(\rho+p)U_{\mid i}~,\\
  \nonumber T^0_{~i}&=&-(\rho+p)(U-B)_{\mid i}~,\\
  T^i_{~j}&=&-p(1+\varepsilon)\delta^i_{~j}-(\rho+p)
        \Sigma_{\mid ij}~,
\end{eqnarray}
here $\rho$ and $p$ are unperturbed density and pressure; $U$ and 
$\Sigma$ determine velocity perturbation and anisotropic shear 
perturbation, 
respectively;
$\epsilon$ and $\varepsilon$ denote the fluctuation for density and 
pressure respectively.

In many gauge-invariant approaches, the gauge-invariant density 
perturbation
was constructed by \cite{bardeen2,kodama1}
\begin{equation}\label{epsilonm}
  \epsilon_m\equiv \epsilon+3(1+w)
        \frac{\dot{a}}{a}(U-B)~,
\end{equation}
where a dot denotes the derivative with respect to the conformal time 
$\tau$; $w=p/\rho$ is the ratio of the pressure to the density of 
background.
$\epsilon_m$ coincides with the density perturbation $\epsilon_{(CTG)}$ 
in the Comoving Time-orthogonal Gauge ($CTG$, in which $U=B=0$), which 
denotes the density perturbation relative to the spacelike hypersurface 
which represents the
matter local rest frame everywhere\cite{bardeen2}. This quantity also 
coincides with the density perturbation $\epsilon_{(SG)}$ in the 
Synchronous Gauge (in which $\Psi=B=0$) for the pressureless matter 
system. In 
other words, $\epsilon_m$ denotes the density perturbation relative to 
the observers everywhere comoving with the matter ({\em not} with the 
unperturbed expanding background of universe because the matter have 
local velocity other than the Hubble flow). So, $\epsilon_m$ has 
physical 
significance only for the small-scale perturbation everywhere, and this 
can be also derived from following facts: The relation between 
$\epsilon_m$ and the general gravitational potential $\phi$ (which was 
correctly constructed in literatures, see Eq.(\ref{potential}) below) 
obeys 
Poisson equation\cite{bardeen2,kodama1}:
\begin{equation}\label{poisson} 
\bigtriangledown^2\phi=-k^2\phi
=4\pi G\rho a^2\epsilon_m~.
\end{equation}
where $k$ is the (comoving) wavenumber of Fourier mode. Poisson 
equation is valid only in the case that the investigated scale is small 
compared to the Hubble radius $1/H$ with $H=\dot{a}/a^2$ being Hubble 
constant \footnote{In the standard big-bang model, the Hubble radius is 
the 
same order as the horizon, but in the inflationary scenario, the 
horizon 
can be much large than the Hubble radius. The largest scale in 
causality is set by the horizon, while the physical perturbation modes 
outside 
the Hubble radius will hardly grow as long as the equation of state for 
the background of universe is unchanged. Notice that the perturbation 
amplitude for all modes will change if the equation of state of the 
background changes, e.g., $\phi$ changes during the transition from 
radiation-domonated era to matter-dominated 
era\cite{mukhanov}.}\cite{lss}. 
Hence, $\epsilon_m$ can not be regarded as the physical density 
perturbation directly for the large-scale (comparable to or larger than 
the 
Hubble radius) perturbation mode in the expanding universe because the 
gravitational potential $\phi$ has been constructed correctly. Another 
fact 
is that the linearized Einstein equation shows that $\epsilon_m$ keeps 
on growing even for the perturbation modes outside both the Hubble 
radius and the horizon
\cite{bardeen2,kodama1}, which also means that $\epsilon_m$ can not 
directly characterize the physical large-scale perturbation modes. Therefore, 
the 
physical density perturbation which comprises both the small-scale and 
the large-scale modes can not be described {\em directly} by 
$\epsilon_m$, though it is a gauge-invariant quantity.

In fact, the large-scale density perturbation has physical significance 
only in the case that the corresponding density perturbation everywhere 
is relative to the unperturbed expanding background of universe. So we 
must construct a proper gauge-invariant variable to describe the 
physical density perturbation {\em directly}, and then confront it with 
the 
large-scale structure of universe.
It is my viewpoint that the gauge-invariant quantity constructed by
\cite{bardeen2,kodama1,mukhanov},
\begin{equation}\label{gidp}
  \epsilon_g\equiv
  \epsilon+\frac{\dot \rho}{\rho}(B-\dot{E})=
\epsilon-3(1+w)\frac{\dot{a}}{a}(B-\dot{E})~.
\end{equation}
is just what we sought for.
$\epsilon_g$ coincides with the density perturbation
$\epsilon_{(NG)}$ in the Newtonian Gauge
($NG$, in which $B=E=0$).
Correspondingly, two gauge-invariant scalar potentials $\phi$ and 
$\psi$,
both of which become the same as the gravitational potential
in the Newtonian limit, are constructed from metric perturbations
\cite{mukhanov}:
\begin{eqnarray}\label{potential}
\nonumber && \phi \equiv \Phi-\frac{\dot{a}}{a}(B-\dot{E})~,\\
      && \psi \equiv \Psi+\frac{1}{a}\frac{d}{d\tau}[(B-\dot{E})a]~.
\end{eqnarray}
The gauge-invariant pressure perturbation $\delta p/p$
and velocity perturbation $v_i$ are constructed as
\begin{equation}\label{p}
  \delta p/p\equiv \varepsilon+\frac{\dot p}{p}
    (B-\dot {E})~,
\end{equation}
\begin{equation}\label{v}
  v_i\equiv (U-\dot{E})_{|i}~.
\end{equation}

The time-time part of the
linearized Einstein equation gives \cite{mukhanov,ma}
\begin{equation}\label{zero}
\bigtriangledown^2\phi-3\frac{\dot{a}}{a}(\frac{\dot{a}}{a}\psi+\dot{\phi})
=-k^2\phi-3\frac{\dot{a}}{a}(\frac{\dot{a}}{a}\psi+\dot{\phi})
=4\pi G\rho a^2\epsilon_g~.
\end{equation}
From this equation we can
see that the relation between the potential $\phi$ and the physical 
density
perturbation $\epsilon_g$ does not
obey Poisson equation any more for the large-scale perturbation modes.
At the same time, it can be deduced from the linearized Einstein 
equation that $\epsilon_g$ hardly grows when the perturbation mode is 
outside 
the Hubble radius ($k<aH$).

Hence, though any gauge can be employed to work for the density 
perturbation,
We should adopt the gauge-invariant variable which characterizes the 
density perturbation relative to the unperturbed expanding background 
of 
universe to compare with the large-scale structure of universe {\em 
directly} (here I do not consider the biasing issue).

For the multi-component system,
$\epsilon_g$, $\delta p/p$ and $v_i$ denote the gauge-invariant
variables of total density perturbation,
total pressure perturbation and total velocity perturbation
respectively, and the corresponding gauge-invariant
variables for $\alpha$-component are constructed in
similar ways \cite{kodama1}:
\begin{equation}
\epsilon_{\alpha g}\equiv
  \epsilon_{\alpha}+\frac{\dot 
\rho_{\alpha}}{\rho_{\alpha}}(B-\dot{E})~,
\end{equation}
\begin{equation}\label{pa}
  (\delta p/p)_{\alpha}
\equiv \varepsilon_{\alpha}+
    \frac{\dot p_{\alpha}}{p_{\alpha}}
    (B-\dot {E})~,
\end{equation}
\begin{equation}\label{va}
  v_{\alpha i}\equiv (U_{\alpha}-\dot{E})_{|i}~.
\end{equation}

From above equations we can see that all these gauge-invariant
perturbation variables conincide with the corresponding ones in the 
Newtonian Gauge
respectively.
We can also adopt the density perturbation defined in this gauge
to compare with the large-scale structure of universe {\em directly}.

Any a real-space fluctuation
$\delta(\bf{x},\tau)$ can be decomposed into
Fourier modes
\begin{equation}\label{delta}
\delta({\bf{x}},\tau)=\frac{1}{(2 \pi)^{\frac{3}{2}}}
\int \delta_{\bf k}(\tau)
e^{i\bf{k\cdot x}}d^3k~,
\end{equation}
with the reality condition $\delta_{\bf k}^{\ast}=\delta_{-\bf{k}}$~.
The power spectrum is defined as the mean square of
the corresponding
Fourier-mode amplitude:
\begin{equation}
P_{\delta}(k,\tau)=<|\delta_{\bf k}(\tau)|^2>\equiv
|\delta(k,\tau)|^2~.
\end{equation}
All the statistical properties of the Gaussian random field can be
determined by the power spectrum completely, and for simplicity
we only consider this case here. At the same time,
we are only interested in the fluctuation 
relative to the unpertured background everywhere which can be
confronted with the large-scale structure of universe directly.
The root mean square (rms) fluctuation
and two-point
correlation function of density on the scale $R$
can be calculated by
the physical power spectrum $P_{\epsilon_g}(k,\tau)\equiv
|\epsilon_g(k,\tau)|^2$ directly
\begin{eqnarray}\label{sigma}
\nonumber
\sigma_{\epsilon_g}(R,\tau)&=&\left[\frac{1}{2\pi^2}
\int_0^{\infty}P_{\epsilon_g}(k,\tau)
W^2(kR)k^3\frac{dk}{k}\right]^{\frac{1}{2}}~,\\
\xi_{\epsilon_g}(R,\tau)&=&\frac{1}{2\pi^2}
\int_0^{\infty}P_{\epsilon_g}(k,\tau)
\frac{\sin kR}{kR}k^3\frac{dk}{k}~,
\end{eqnarray}
where, $W(x)=\frac{3}{x}j_1(x)$ 
is the window function
which truncates the contribution
of the power spectrum of large wavenumbers to the rms fluctuation
(here $j_1$ is the first-order spherical Bessel function).
When calculating $\sigma_{\epsilon_g}(R,\tau)$ or 
$\xi_{\epsilon_g}(R,\tau)$, we need to
integrate the power spectrum over all wavenumbers, not merely the modes 
within the Hubble radius (see Eq.(\ref{sigma})).
Because the spectrum $P_{\epsilon_m}(k,\tau)\equiv
|\epsilon_m(k,\tau)|^2$
differs dramatically from $P_{\epsilon_g}(k,\tau)$ for
small wavenumbers (e.g., $k\le aH$),
generally we can not replace $P_{\epsilon_g}(k,\tau)$
with $P_{\epsilon_m}(k,\tau)$ in Eq.(\ref{sigma}) directly to obtain 
the
rms fluctuation and correlation function of
the density.

The temperature fluctuation of the
cosmic microwave background (CMB)
can be characterized by the angular power spectrum $C_l$.
Based on the line-of-sight-integral method developed by
Seljak and Zaldarriaga\cite{seljak}, we have:
\begin{eqnarray}\label{cl}
  C_l=\frac{1}{2\pi}\int_0^{\infty}\frac{dk}{k}
        k^3\left[\int_0^{\tau_0}\Theta(k,\tau)
        j_l(k\tau_0-k\tau)d\tau\right]^2~,
\end{eqnarray}
with
\begin{eqnarray}\label{effects}
\nonumber  \Theta(k,\tau) &=&
        g\left(\frac{1}{4}\epsilon_{\gamma}+\phi+
    \frac{\dot{v}_{b}}{k}+
\frac{\dot{g}}{g}\frac{v_{b}}{k}\right)+e^{-\kappa}(\dot{\phi}+\dot{\psi})\\
&& +g\left(\frac{\prod}{4}+\frac{3\ddot{\prod}}{4k^2}\right)
        +\dot{g}
        \frac{3\dot{\prod}}{4k^2}
        +\ddot{g}\frac{3\prod}{4k^2}~,
\end{eqnarray}
where $j_l$ is the spherical Bessel function;
$\epsilon_{\gamma}$ is the photon density
perturbation; $\prod$ is a gauge-invariant quantity denoting the 
effect of
polarization and anisotropic stress of photons;
$v_{b}$ denotes the velocity
perturbation of baryons;
$\tau_0$ denotes present time;
$g$ is the visibility function and $\kappa$
is the total optical depth at time $\tau$.

Now, let us take the standard cold dark matter (SCDM) model as an 
example to illustrate the prediction differences between the 
conventional 
treatment and my viewpoint. The SCDM model (the cold dark matter 
density 
parameter $\Omega_c=0.95$, the baryon density parameter 
$\Omega_b=0.05$, 
$H_0=100h ~km~s^{-1}~Mpc^{-1}$ with $h=0.5$) assumes that
the primordial density spectrum in the $SG$ frame
$P_{\epsilon(SG)}(k,\tau_i)$
is the Harrizon-Zel'dovich (HZ) spectrum:
\begin{equation}\label{hz}
  P_{\epsilon(SG)}(k,\tau_i)\equiv \mid 
\epsilon_{(SG)}(k,\tau_i)\mid^2=Ak~,
\end{equation}
here $A$ is a constant and $\tau_i$ is the primordial time.
In the conventional treatment,
the density perturbation $\epsilon_{(SG)}$ in the $SG$ frame was 
directly
regarded as the physical density perturbation,
so the rms fluctuation of the density on the scale of
$r_8=8h^{-1}~Mpc$
was directly corresponded to $\sigma_{\epsilon (SG)}(r_8,\tau_0)$,
which is related to $P_{\epsilon(SG)}(k,\tau_0)\equiv \mid 
\epsilon_{(SG)}(k,\tau_0)\mid^2$ by
\begin{eqnarray}\label{rms8sg}
  \sigma_{\epsilon 
(SG)}(r_8,\tau_0)=\left[\frac{1}{2\pi^2}\int_0^{\infty}
          P_{\epsilon(SG)}(k,\tau_0)W^2(kr_8)
              k^3\frac{dk}{k}\right]^{\frac{1}{2}}~.
\end{eqnarray}

In the foregoing discussions
we have shown that this picture is wrong because 
$\epsilon_{(SG)}(k,\tau)$ (or its spectrum 
$P_{\epsilon(SG)}(k,\tau)\equiv \mid \epsilon_{(SG)}(k,\tau) \mid^2$)
can not characterize the physical density fluctuation {\em directly}.
The physical density spectrum corresponding to
the assumption of the SCDM model on the primordial density spectrum
should be
\begin{equation}\label{scdmp}
  P_{\epsilon_g}(k,\tau)=
P_{\epsilon_g}(k,\tau_i)T^2_{\epsilon_g}(k,\tau)\simeq
\left(\frac{3a^2_iH^2_i}{k^2}\right)^2 Ak T^2_{\epsilon_g}(k,\tau)~,
\end{equation}
where, the subscript ``$_i$'' denotes the primordial time;
$P_{\epsilon_g}(k,\tau_i)\simeq (3a^2_iH^2_i/k^2)^2 Ak $
is the physical primordial spectrum corresponding to the HZ spectrum
in the $SG$ frame (notice that all the interesting 
modes are far outside the Hubble radius at time $\tau_i$, i.e., $k<<a_iH_i$), 
and $T_{\epsilon_g}(k,\tau)\equiv \frac{\epsilon_g (k,\tau)}{\epsilon_g 
(k,\tau_i)}$
is the transfer function of
physical density spectrum due to the evolution of
density perturbation from the primordial time
$\tau_i$ to the time $\tau$.

Fig.1 shows the physical density spectrum $P_{\epsilon_g}(k,\tau)$
of the SCDM model
at the redshift $z=0$ and $z=10$,
with the corresponding
$P_{\epsilon(SG)}(k,\tau)$. 
It can be seen that $P_{\epsilon(SG)}(k,\tau)$
can approximate the physical spectrum only for the large wavenumbers
(small-scale modes),
and differs significantly from the latter for the wavenumbers outside 
the Hubble radius.
By the way, $P_{\epsilon(SG)}(k,\tau)$ differs hardly from the
gauge-invariant power spectrum $P_{\epsilon_m}(k,\tau)$ 
for the SCDM model in which the universe
was dominated by the pressureless matter at very high redshift.
In the conventional treatment, only the power spectrum of the 
wavenumbers inside the Hubble radius is taken seriously, and the part 
outside
the Hubble radius is thought to have no physical significance.
This viewpoint is not correct:
When a real-space fluctuation is decomposed into
Fourier modes,
all these modes will have physical significance
whether or not they are inside the Hubble radius (see 
Eq.(\ref{delta})).
Only the {\em whole} power spectrum
can determine a real-space Gaussian fluctuation completely.
In fact, the power spectrum of the wavenumbers outside the Hubble 
radius
has also been included
in the current calculation of the density fluctuations
(e.g., see Eq.(\ref{rms8sg}))
and the angular power spectrum
$C_l$ for CMB (see Eq.(\ref{cl})).
So, the conventional treatment is inconsistent upon this point.
More important, the temperature fluctuation detected by the COBE 
satellite
(e.g., $C_2$) is dominated by the power spectrum of the wavenumbers
outside the Hubble radius at the recombination era!
On the other hand, inflationary scenario has provided a well-known 
mechanism
of the non-vanishing power spectrum for the wavenumbers outside the 
Hubble radius.

Fig.1 also reveals another important difference
between the physical spectrum and $P_{\epsilon(SG)}(k,\tau)$:
the shape of the physical spectrum changes
even for the pressureless matter system,
because the physical density perturbation modes far outside the Hubble 
radius
can not grow due to the causality;
while the shape of $P_{\epsilon(SG)}(k,\tau)$ for the SCDM model hardly 
changes,
because the growth rate of $\epsilon_{(SG)}(k,\tau)$ is
independent of wavenumber $k$ for the presureless system.

Now we re-check some predictions of the SCDM model.
From Eqs.(\ref{sigma}) and (\ref{scdmp}) we find that
the rms fluctuation $\sigma_{\epsilon_g}(R,\tau)$
and the correlation $\xi_{\epsilon_g}(R,\tau)$ of density
are both divergent on any scale $R$ at any time $\tau$,
so do those of the potential $\phi$, {\em if} the physical density
spectrum has the HZ shape exactly in the limit of $k\rightarrow 0$ in 
the $SG$ frame. In fact, these unreasonable predictions caused by the 
divergent spectrum
have also been reflected in the calculations of the temperature
fluctuation of CMB, though it is very implicit. 
From the combination of Eqs.(\ref{cl}), (\ref{effects}) and 
(\ref{scdmp}) we can obtain that $C_0$ is divergent too. This means 
that
the temperature fluctuation of CMB is infinite in any direction.
Because people are only interested in the angular distribution of the
temperature anisotropy of CMB (e.g., $C_l$ for
$l\ge 1$), this unreasonable prediction of the SCDM model has been 
ignored so far.

All these unreasonable predictions are due to the fact that the power
spectrum was not cut off effectively.
Any physical power spectrum should become negligible as
wavenumber $k$ tends to zero due to the causality.
So, the primordial spectrum should be cut off effectively at
a minimum wavenumber $k_{min}$, e.g., $a/k_{min}$ is at most
the same order as the horizon (notice that the horizon may be much 
larger
than the Hubble radius). After doing so, we can see that all 
calculation results are convergent and dependent on $k_{min}$.

The key point lies in how to determine the effective
minimum wavenumber $k_{min}$ for the power spectrum.
If $k_{min}$ is inside the Hubble radius at any time, there will not be 
so many problems to discuss here. However, $k_{min}$ can be far outside 
the Hubble radius (for example, the quantum fluctuation in the early 
universe can be stretched far outside the Hubble radius by inflation 
process),
otherwise we can not explain the temperature fluctuation detected by
the COBE satellite, which showed that these fluctuation
modes are outside the Hubble radius at the recombination era.

In conclusion, the power spectrums differ dramatically for the modes
outside the Hubble radius in different gauges, however, the calculation 
of the rms density fluctuation needs to integrate the power
spectrum for all perturbation modes
(see Eq.(\ref{sigma})),
not merely the modes within the Hubble radius, so do the calculations 
of the
two-point correlation function, the
abundance of the galaxies or clusters by Press-Schechter formula,
and the angular power spectrum $C_l$ for CMB, etc. 
In the above demonstrations 
we have shown that these quantities should be calculated directly by
the gauge-invariant spectrum $P_{\epsilon_g}(k,\tau)$,
or the spectrum $P_{\epsilon(NG)}(k,\tau) \equiv \mid 
\epsilon_{(NG)}(k,\tau)\mid^2$ in the Newtonian Gauge
which coincides with $P_{\epsilon_g}(k,\tau)$,
but not directly by the spectrum $P_{\epsilon(SG)}(k,\tau)$
in the Synchronous Gauge.
On the other hand, the density fluctuations and the CMB anisotropies
are dependent on the cut-off wavenumber $k_{min}$ of
the physical power spectrum.
$k_{min}$ is not set by the Hubble radius simply, but by the
physical process in early universe, such as inflation. 
In the inflationary picture there generally exists a non-vanishing 
power spectrum for the modes outside the Hubble radius, i.e., $k_{min}$ 
can 
be far outside the Hubble radius!
These conclusions will change dramatically
the scenarios of the current cosmological models on
the formation of the large-scale structure of universe.

~\\
\indent I am thankful to R. Brandenberger for reading manuscript 
and
giving valuable comments. I also thank
E. Bertschinger for using the COSMICS,
and U. Seljak and M. Zaldarriaga
for using the CMBFAST. This work was supported in part
by the National Natural Science Foundation of China under Grant 
No.10047004.

\newpage
\begin{Large}
{\bf Captions for Figure}
\end{Large}
\vskip 5mm
Fig.1 The physical density spectrum for the SCDM model (denoted as the
solid lines), with the corresponding spectrum in the Synchronous Gauge
(denoted as the dashed lines).
Since the physical rms fluctuation
$\sigma_{\epsilon_g}(r_8,\tau_0)$ (see Eq.(\ref{sigma}))
is divergent for the SCDM model, for simplicity the
normalization for the density spectrum
is chosen so that $\sigma_{\epsilon(SG)}(r_8,\tau_0)=1$
(see Eq.(\ref{rms8sg})) instead;
The upper and the lower lines correspond to the cases of $z=0$ and 
$z=10$,
respectively.

\newpage

\begin{figure}
\epsfxsize=6 in \epsfbox{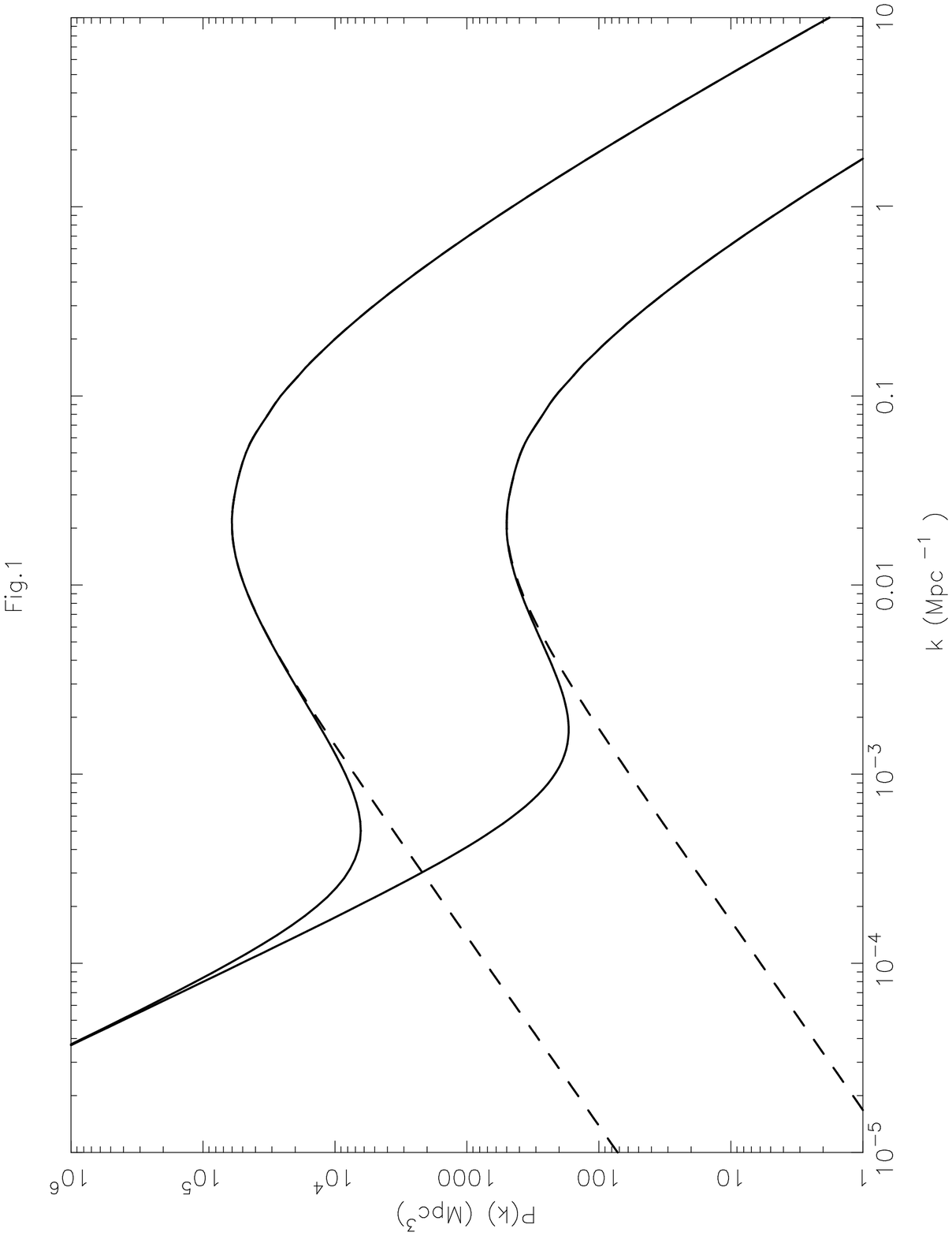}
\end{figure}

\end{document}